\documentclass[english,aps,prx,reprint,longbibliography]{revtex4-2}
\usepackage{amssymb}
\usepackage{amsmath}
\usepackage[version=4]{mhchem} 
\usepackage{siunitx}[=v2]
\usepackage{graphicx}
\usepackage{natbib}
\usepackage{epstopdf}
\usepackage{color}
\usepackage{braket}
\usepackage{physics}
\usepackage{mathrsfs}
\usepackage{upgreek}
\usepackage{todonotes}
\usepackage[colorlinks=true, allcolors=blue, urlcolor=black]{hyperref}
\usepackage{soul}
\usepackage{lipsum}
\usepackage{afterpage}
\usepackage[super]{nth}
\usepackage[modulo]{lineno}

\newcommand{\figref}[1]{Fig.~\ref{#1}}
\newcommand{\J}[1]{$J_{\rm #1}$}

\begin{document}
\title{Two-dimensional Si spin qubit arrays with multilevel interconnects}

\author{Sieu D. Ha}
\affiliation{HRL Laboratories, LLC, 3011 Malibu Canyon Road, Malibu, California 90265, USA}
\author{Edwin Acuna}
\email{Corresponding author: eacuna@hrl.com}
\affiliation{HRL Laboratories, LLC, 3011 Malibu Canyon Road, Malibu, California 90265, USA}
\author{Kate Raach}
\affiliation{HRL Laboratories, LLC, 3011 Malibu Canyon Road, Malibu, California 90265, USA}
\author{Zachery T. Bloom}
\affiliation{HRL Laboratories, LLC, 3011 Malibu Canyon Road, Malibu, California 90265, USA}
\author{Teresa L. Brecht}
\affiliation{HRL Laboratories, LLC, 3011 Malibu Canyon Road, Malibu, California 90265, USA}
\author{James M. Chappell}
\affiliation{HRL Laboratories, LLC, 3011 Malibu Canyon Road, Malibu, California 90265, USA}
\author{Maxwell D. Choi}
\affiliation{HRL Laboratories, LLC, 3011 Malibu Canyon Road, Malibu, California 90265, USA}
\author{Justin E. Christensen}
\affiliation{HRL Laboratories, LLC, 3011 Malibu Canyon Road, Malibu, California 90265, USA}
\author{Ian T. Counts}
\affiliation{HRL Laboratories, LLC, 3011 Malibu Canyon Road, Malibu, California 90265, USA}
\author{Dominic Daprano}
\affiliation{HRL Laboratories, LLC, 3011 Malibu Canyon Road, Malibu, California 90265, USA}
\author{J. P. Dodson}
\affiliation{HRL Laboratories, LLC, 3011 Malibu Canyon Road, Malibu, California 90265, USA}
\author{Kevin Eng}
\affiliation{HRL Laboratories, LLC, 3011 Malibu Canyon Road, Malibu, California 90265, USA}
\author{David J. Fialkow}
\affiliation{HRL Laboratories, LLC, 3011 Malibu Canyon Road, Malibu, California 90265, USA}
\author{Christina A. C. Garcia}
\affiliation{HRL Laboratories, LLC, 3011 Malibu Canyon Road, Malibu, California 90265, USA}
\author{Wonill Ha}
\affiliation{HRL Laboratories, LLC, 3011 Malibu Canyon Road, Malibu, California 90265, USA}
\author{Thomas R. B. Harris}
\affiliation{HRL Laboratories, LLC, 3011 Malibu Canyon Road, Malibu, California 90265, USA}
\author{nathan holman}
\affiliation{HRL Laboratories, LLC, 3011 Malibu Canyon Road, Malibu, California 90265, USA}
\author{Isaac Khalaf}
\affiliation{HRL Laboratories, LLC, 3011 Malibu Canyon Road, Malibu, California 90265, USA}
\author{Justine W. Matten}
\affiliation{HRL Laboratories, LLC, 3011 Malibu Canyon Road, Malibu, California 90265, USA}
\author{Christi A. Peterson}
\affiliation{HRL Laboratories, LLC, 3011 Malibu Canyon Road, Malibu, California 90265, USA}
\author{Clifford E. Plesha}
\affiliation{HRL Laboratories, LLC, 3011 Malibu Canyon Road, Malibu, California 90265, USA}
\author{Matthew J. Ruiz}
\affiliation{HRL Laboratories, LLC, 3011 Malibu Canyon Road, Malibu, California 90265, USA}
\author{Aaron Smith}
\affiliation{HRL Laboratories, LLC, 3011 Malibu Canyon Road, Malibu, California 90265, USA}
\author{Bryan J. Thomas}
\affiliation{HRL Laboratories, LLC, 3011 Malibu Canyon Road, Malibu, California 90265, USA}
\author{Samuel J. Whiteley}
\affiliation{HRL Laboratories, LLC, 3011 Malibu Canyon Road, Malibu, California 90265, USA}
\author{Thaddeus D. Ladd}
\affiliation{HRL Laboratories, LLC, 3011 Malibu Canyon Road, Malibu, California 90265, USA}
\author{Michael P. Jura}
\affiliation{HRL Laboratories, LLC, 3011 Malibu Canyon Road, Malibu, California 90265, USA}
\author{Matthew T. Rakher}
\affiliation{HRL Laboratories, LLC, 3011 Malibu Canyon Road, Malibu, California 90265, USA}
\author{Matthew G. Borselli}
\affiliation{HRL Laboratories, LLC, 3011 Malibu Canyon Road, Malibu, California 90265, USA}

\begin{abstract} 
The promise of quantum computation is contingent upon physical qubits with both low gate error rate and broad scalability.  Silicon-based spins are a leading qubit platform, but demonstrations to date have not utilized fabrication processes capable of extending arrays in two dimensions while maintaining complete control of individual spins.  Here, we implement an interconnect process, common in semiconductor manufacturing, with multiple back-end-of-line layers to show an extendable two-dimensional array of spins with fully controllable nearest-neighbor exchange interactions.  In a device using three interconnect layers, we encode exchange-only qubits and achieve average single-qubit gate fidelities consistent with single-layer devices, including fidelities greater than 99.9\%, as measured by blind randomized benchmarking.  Moreover, with spin connectivity in two dimensions, we show that both linear and right-angle exchange-only qubits with high performance can be formed, enabling qubit array reconfigurability in the presence of defects.  This extendable device platform demonstrates that industrial manufacturing techniques can be leveraged for scalable spin qubit technologies.

\end{abstract}

\maketitle

Silicon quantum dot spin qubits are a leading device technology for quantum information processing due to advantages in coherence and the ability to leverage industrial semiconductor manufacturing \cite{BurkardRMP,stuyck2024CMOS}.  Scaling to larger arrays of dots while maintaining independent control of each spin, however, requires an interconnect fabrication process in which signal routing lines are in planes topographically separate from gate electrodes, as in integrated circuits.  There are growing efforts to fabricate qubits in industrial plants, but such processes to date have used in-line routing at the gate level, which has inherent challenges with interior gate connectivity in scaled dot arrays \cite{10413763,Zwerver2022,Elsayed2024,Philips2022}.
Crossbar structures without interconnects can be used to scale arrays of dots but they require common qubit control within the array \cite{2209.06609}, which places challenging bounds on yield and uniformity. 
As a requisite step towards extending quantum dot arrays, we have previously fabricated Si/SiGe qubit devices utilizing one interconnect layer for signal routing and a front-end gate process known as SLEDGE (Single-Layer Etch-Defined Gate Electrode) to encode highly performant exchange-only (EO) qubits in linear and triangular arrangements \cite{doi:10.1021/acs.nanolett.1c03026, HRL_2QEO,Acuna2024}.

\begin{figure}[t!]
	\includegraphics[width=1.0\columnwidth]{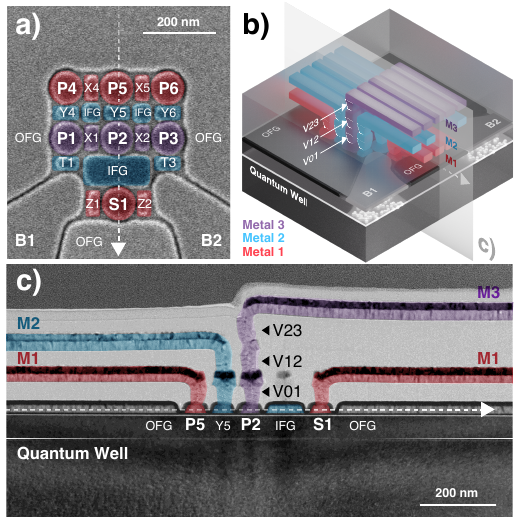}
	\caption{
		(a) Gate-level SEM of the $2\!\times\!3$ plunger gate array with colored overlays denoting
		the BEOL layer used for signal routing, as shown in (b,c). 
		(b) Isometric render of the quantum dot device, including the Si/SiGe heterostructure, vias, and BEOL layers.
		(c) A false-colored cross-sectional TEM of the device, taken along the white arrow in (a), 
		showing all three BEOL layers and vias used for routing electrical signals to gates.  
		Via and BEOL metal layers are denoted $\mathrm{V}xy$ and $\mathrm{M}x$ (or Metal $x$), where $\mathrm{V}xy$ connects $\mathrm{M}x$ and $\mathrm{M}y$,
		starting with V01 between gate metal and M1.
		Note that the partially visible via to the IFG is due to finite TEM lamella thickness, but it is out of plane of the intended cross-section.}
	\label{fig:device}
\end{figure}

\begin{figure}[t]
	\includegraphics[width=\columnwidth]{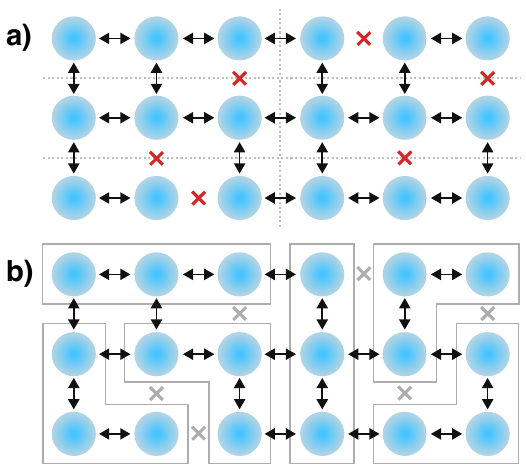}
	\caption{
		(a) A hypothetical 2D array of spins showing available exchange interactions (arrows) and nonfunctional interactions (red crosses). The nonfunctional interactions, which may be intended by design or due to defects, reduce the quantity of available EO qubits when only linear TQD configurations are considered.
		(b) Using both linear and elbow arrangements of TQDs to encode EO qubits, qubit instantiations within the array can be configured to optimize qubit quantity and connectivity.
		}
	\label{fig:reconfig}
\end{figure}

Here, we scale the SLEDGE process to multiple back-end-of-line (BEOL) layers and demonstrate an extendable two-dimensional (2D) spin qubit array platform. The device presented here utilizes five rows of gate electrodes: two rows for qubit formation, one row for charge sensing, one row for inter-qubit exchange, and one row for charge loading.  We show that there is no measurable loss in EO qubit performance with the addition of multiple BEOL layers, as quantified by single-qubit blind randomized benchmarking (BRB) performance.  Moreover, with full nearest-neighbor exchange connectivity we can define three-spin EO qubits in both linear and elbow arrangements, allowing for the flexibility to route qubits around potential defects and optimize total array performance.  The capability to fabricate extendable quantum dot devices enables the investigation of large 2D exchange-coupled spin qubit arrays.

The SLEDGE platform allows for a highly flexible arrangement of gates with a separate BEOL process \cite{Acuna2024}. It is also compatible with non-EO qubit implementations and control.  
In \figref{fig:device}(a), we show a scanning electron microscope (SEM) image of a device similar to the one tested, which is designed nominally for EO operation.  Although the array is approximately a $5\!\times\!5$ grid of all gates, we denote such a device a ``$2\!\times\!3$'' based on two rows of three plunger (P) gates each.  Electrons confined under the two rows of P gates are coupled via the exchange interaction modulated by gates denoted X and Y, with X gates for intra-row exchange and Y gates for inter-row exchange.  Electrons are loaded into the device through bath (B) and tunnel (T/Z) gates, and inner/outer field gates (IFG/OFG) ensure the depletion of electrons near the quantum dot array. The S1 gate is used to form a quantum dot for charge sensing \cite{PRXQuantum.3.010352}. Gate connectivity with multiple BEOL layers is illustrated in \figref{fig:device}(b), with the topmost and bottommost rows of gates contacted with the first BEOL routing layer and inner rows utilizing higher BEOL layers.
A cross-sectional transmission electron microscope (TEM) image of the $2\!\times\!3$ device along the white arrow shown in \figref{fig:device}(a) is given in \figref{fig:device}(c) and shows all three BEOL metal levels and vias. 
Generally $N$ rows of plunger gates require $N+1$ BEOL layers because of the additional row of charge sensors, assuming equal BEOL and gate pitch.  

The SLEDGE fabrication process is detailed in \cite{doi:10.1021/acs.nanolett.1c03026} for a linear six-dot, or ``$1\!\times\!6$,'' array.  To summarize, gates are fabricated by depositing a blanket \ce{Al2O3}/\ce{HfO2}/TiN film stack by atomic layer deposition (ALD), followed by subtractively etching TiN to define gates.  A \ce{SiO2} interlayer dielectric (ILD) is deposited on top of the gates and planarized with chemical-mechanical polishing (CMP).  Vias are patterned and etched through the ILD and are then filled by ALD TiN.  Routing lines can be defined by a variety of integration schemes (e.g. subtractive, single or dual damascene), but here we use a subtractive approach with TiN/W metal lines formed by TiN via fill overburden and a blanket W deposition.  Extending the process to multiple BEOL layers and P gate rows is straightforwardly accomplished by repeating the steps from ILD deposition through TiN/W metal patterning.  Note that Ohmic and isolation implants are performed prior to gate stack deposition, and all nanoscale features (gate, via, BEOL metal) are defined using electron beam (e-beam) lithography.  The Si/SiGe heterostructure here uses a \SI{3}{\nm} thick Si quantum well with an isotopic concentration of 800 ppm $^{29}$Si and isotopically natural Si and Ge in the adjoining SiGe layers. 

Exchange-only qubits are instantiated using a triple-quantum-dot (TQD) decoherence-free subsystem (DFS) encoding \cite{DiVinceEO}.  The three-spin DFS encoding consists of a singlet-triplet pair [denoted ($a$,$b$) for spins $a$ and $b$] and one additional spin [arriving at $(a,\!b)c$ for additional spin $c$], with the singlet-triplet pair usually considered in the first two spins for three spins in a row [i.e. (1,2)3, where the indices refer to the indices of three consecutive corresponding P-gates]. 
There are, however, three nontrivial even permutations of spin assignments for such a row [i.e. (1,2)3, (2,3)1, and (3,1)2] within a TQD for the EO DFS encoding. Of these permutations there are two with singlet-triplet spin pairs connected by a single exchange axis [(1,2)3 and (2,3)1]. These two configurations enable $\sigma^{z}$ and  $\sigma^{n}$ rotations without additional spin swaps and are the permutations we consider in this work. Thus our qubit control Hamiltonian takes the familiar form
$H_{\mathrm{X}i\mathrm{X}j} = J_{{\rm X}i}\sigma^{z} + J_{{\rm X}j}\sigma^{n}$, where $\sigma^{n} = -(\sqrt{3}\sigma^{x}  + \sigma^{z})/2$. Here, we denote our qubits as ${{\rm X}i{\rm X}j}$ for the two exchange gates in a given TQD,
where exchange energies $J_{{\rm X}i}$ and $J_{{\rm X}j}$ modulate the $\sigma^{z}$ and $\sigma^{n}$ rotations, respectively. 

Although EO qubits are often considered in linear TQD arrangements, they can also be formed in both linear and right-angle elbow configurations within a 2D rectangular lattice.  This is advantageous because if a small subset of plunger or exchange gates in an array do not function properly (red crosses in \figref{fig:reconfig}(a)), it remains possible to form coupled EO qubits around the defects, as shown in \figref{fig:reconfig}(b).  Such reconfigurability can mitigate realistic yield limitations due to fabrication, SiGe alloy disorder, or other performance non-uniformity.  
In an $n\!\times\!m$ rectangular array of quantum dots with all vertical and horizontal nearest-neighbor interactions, as in \figref{fig:device}(a), it can be shown that there are $6(n-1)(m-1)-2$ arrangements of single TQDs.
Thus in our $2\!\times\!3$ device, there are 10 TQDs and 20 EO DFS qubit assignments available, as well as six possible assignments of EO qubit pairs.

\begin{figure}[t]
	\includegraphics[width=\columnwidth]{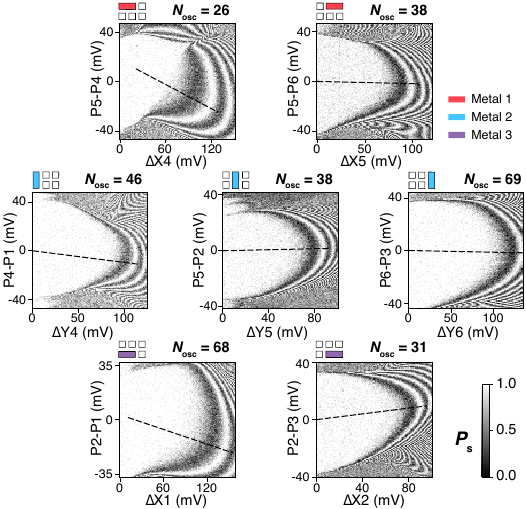}
	\caption{
		Exchange ``fingerprint'' plots for each of the seven possible exchange interactions in a $2\!\times\!3$ device. 
		The colored diagrams illustrate the exchange interaction site under investigation, and the color denotes the BEOL layer used to contact the barrier gate, 
		with the colors matching those in \figref{fig:device}.
		The black dashed lines denote the symmetric operating vector along which exchange rotations are calibrated.
		Calibrated spin swaps are performed to move one of the spin singlet electrons
		to different parts of the array in order to observe exchange oscillations (see text). 
		The singlet return probability is measured at the P2-P3 DQD. 
		There is no clear correlation between BEOL layer used to contact an exchange gate and its $N_{\rm osc}$ parameter. 
		}
	\label{fig:fingerprints}
\end{figure}

The EO encoding is straightforward to implement in SLEDGE-based devices because it requires only exchange-based control, modulated using baseband voltage pulses, for both state preparation and measurement (SPAM) as well as qubit manipulations \cite{Andrews2019,HRL_2QEO}.
We utilize Pauli spin blockade for readout and initialization of EO qubits \cite{PhysRevApplied.12.014026, PRXQuantum.3.010352}. 
The encoded $|0\rangle$ state is defined as a two-spin singlet and a third electron left in an arbitrary spin state. 
For the demonstrations in this work, there are six electrons loaded under the array of P gates, with a single electron occupying each quantum dot. 
Here, we use the P2-P3 double-quantum-dot (DQD) to initialize spin singlets and to discriminate between singlet and triplet spin states, though other DQDs in the array are viable options. 
To initialize an arbitrary EO TQD qubit in the 2D array into encoded $|0\rangle$ as well as to measure its spin state, we utilize exchange-based calibrated spin swaps to coherently transfer two-spin states back and forth between the P2-P3 DQD and any arbitrary DQD \cite{Kandel2019}. 
The $2\!\times\!3$ array could additionally support simultaneous initialization of two qubits, 
but it is not required for the single-qubit demonstrations in this work; 
readout of two EO qubits in the present device would require the readout to be performed sequentially. 
Future work will involve demonstrations and characterization of multiqubit initialization in 2D arrays.

Exchange-only operation means that, up to the mathematical assignment of qubits, 
all operation amounts to calibrated exchange operations at each axis.  
Hence, we effectively calibrate arbitrary qubits by arbitrary exchange axes, and then assignment of qubits is straightforward. 
Exchange axes can be calibrated by first calibrating exchange near the SPAM location 
and then swapping initialized spin states to locations of interest and back.  
In particular, as the two-spin singlet is initialized in the P2-P3 DQD [denoted as (2,3)$c$], 
\J{X2} natively causes $\sigma^{z}$ rotations, the observation of which is dependent upon the calibration of $\sigma^{n}$ rotations. 
Energies \J{X1}, \J{Y5}, and \J{Y6} can all readily be used to modulate and observe $\sigma^{n}$ interactions [where $c$ would be the spins in dots P1, P5, or P6, respectively] and are thus calibrated first; 
the ``fingerprint'' plots for these and all other axes are shown in \figref{fig:fingerprints}. 
From these exchange fingerprints, and utilizing the symmetric mode of operation ~\cite{PhysRevLett.116.110402,PhysRevLett.116.116801}, 
we calibrate the \J{X1}, \J{Y5}, and \J{Y6} exchange energies using techniques developed in prior work \cite{Andrews2019}. 
We then perform calibrated spin swaps to move one of the singlet spins, which initially reside in the P2-P3 DQD, 
to measure and calibrate exchange at the remaining X and Y interaction sites. 
As an example, a calibrated spin swap induced by \J{X1} would place the singlet spins in dots P1 and P3 [denoting this permutation (1,3)$c$], and interactions between one of these two spins and any other adjacent spin would induce observable $\sigma^{n}$ interactions.

Utilizing calibrated spin swaps,  
we characterize the 2D quantum dot array 
while maintaining single-electron occupancy of all the quantum dots \cite{Kandel2019}.
For each exchange interaction site, we measure the number of exchange oscillations at the $1/e$ decay point, $N_{\rm osc}$. 
This parameter quantifies the impact of charge noise and its manifestation as exchange noise \cite{HRL_2QEO}. 
The values of $N_{\rm osc}$ for each exchange axis are shown in \figref{fig:fingerprints}, with $J \approx \SI{100}{MHz}$. 
We observe no clear correlation between the BEOL layer used to connect to an X or Y barrier gate and the resulting $N_{\rm osc}$ parameterization, and we observe exchange coherence in our multilayer BEOL device consistent with demonstrations in devices utilizing a single interconnect layer \cite{HRL_2QEO,Acuna2024}. 

\begin{figure*}[t!]
	\includegraphics[width=2.0\columnwidth]{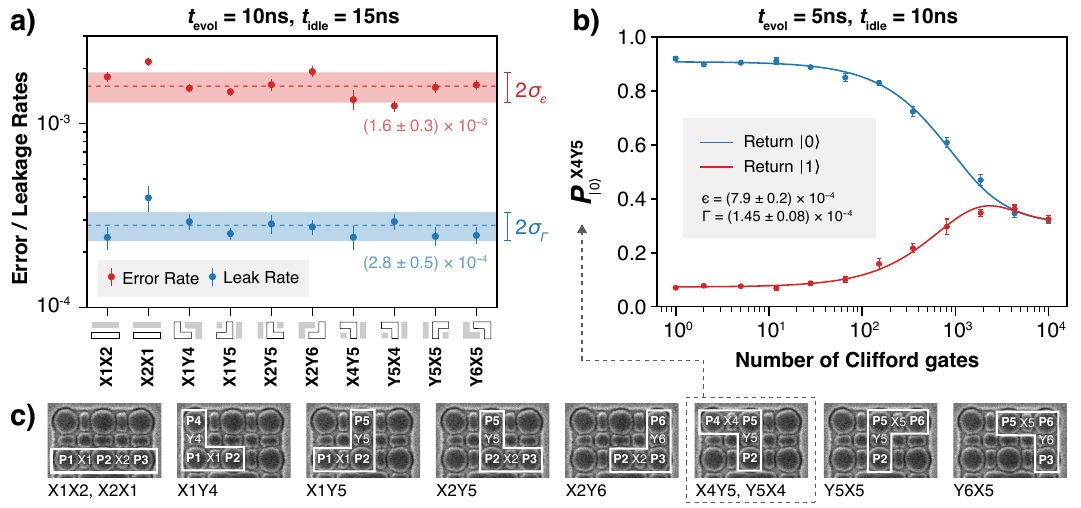}
	\caption{
		(a) Exchange-only qubit characterization in a $2\!\times\!3$ device. 
		Eight unique pairwise combinations of exchange axes (i.e., eight TQDs), as outlined in (c), 
		are used to encode 10 EO qubits. 
		Singlet initialization and qubit state readout are performed via Pauli spin blockade at the P2-P3 DQD, 
		and the qubits are initialized using calibrated spin swaps to place the singlet spins at the $\sigma^{z}$ sites. 
		Qubit fidelities are characterized using BRB~\cite{Andrews2019}. 
		With a $B=\SI{1}{mT}$ applied magnetic field, the average error rate across these qubits is $(1.6 \pm 0.3) \times 10^{-3}$ and the 
		average leakage rate is $(2.8 \pm 0.5) \times 10^{-4}$; 
		the average rates are shown as dashed lines and the shaded regions are twice the standard deviations of the respective distributions.
		(b)~One of the five BRB experiments of qubit X4Y5 using durations $t_{\rm pulse} = \SI{5}{\ns}$ and $t_{\rm idle} = \SI{10}{\ns}$, 
		resulting in error and leakage rates $\epsilon = (7.9 \pm 0.2) \!\times\! 10^{-4}$ and $\Gamma = (1.45 \pm 0.08) \!\times\! 10^{-4}$, respectively. 
		(c)~The eight TQDs used to encode the 10 EO qubits characterized in (a). 
		}
	\label{fig:tetris}
\end{figure*}

We evaluate the performance of various qubits encoded in this 2D array by utilizing multiple applications of calibrated spin swaps to coherently transfer the spin singlet from P2-P3 to any other DQD in the spin array.  For the demonstrations in this work, we characterize 10 of the 20 available EO qubit assignments in the $2\!\times\!3$ device over eight TQD configurations, with the set of characterized qubits spanning each exchange interaction axis at least once. 
The eight TQDs used to encode the 10 EO qubits (in two of the TQDs we characterized two spin permutations instead of one) are shown in \figref{fig:tetris}(c). 
Utilizing the BRB technique~\cite{Andrews2019}, we characterize both the qubit leakage rate and the total error, 
with the BRB results of the 10 EO qubit permutations shown in \figref{fig:tetris}(a). 
Using pulse and idle durations of $t_{\rm pulse} = \SI{10}{\ns}$ and $t_{\rm idle} = \SI{15}{\ns}$, the average single-qubit error and leakage rates in this device are $\epsilon_{\rm avg} = (1.6 \pm 0.3) \!\times\! 10^{-3}$ and $\Gamma_{\rm avg} = (2.8 \pm 0.5) \!\times\! 10^{-4}$. 
We observe a narrow distribution in qubit performance over the various TQDs, noting we have measurements for only a subset of the available single-qubit encodings in this 2D array. 
The EO single-qubit fidelities in this multilayer BEOL device are comparable with recent demonstrations in the SLEDGE platform using single-layer BEOL processes \cite{HRL_2QEO,Acuna2024,doi:10.1021/acs.nanolett.1c03026}.
We can improve BRB performance by decreasing the average gate time, consistent with reducing hyperfine error contributions to the gate error as discussed in \cite{HRL_2QEO}. 
For qubit X4Y5, a weighted average of five repetitions of BRB using durations $t_{\rm pulse} = \SI{5}{\ns}$ and $t_{\rm idle} = \SI{10}{\ns}$ results in 
total error rate $\epsilon = (7.9 \pm 0.2) \!\times\! 10^{-4}$ and qubit leakage rate $\Gamma = (1.45 \pm 0.08) \!\times\! 10^{-4}$, 
reaching an average single-qubit gate fidelity greater than 99.9\% (\figref{fig:tetris}(b) shows one of the five repetitions).
The spin array here, with all nearest-neighbor exchange couplings calibrated, is capable of two-qubit EO operations, enabling studies similar to Ref. \cite{HRL_2QEO}.

While the SLEDGE-based 2D spin qubit array we have demonstrated here is technically arbitrarily extendable in two dimensions by adding more BEOL layers, the ultimate array size is effectively constrained by process capabilities, finite material properties, and electrostatic limitations.   Overlay is a particular process concern, with \SI{140}{\nm} P-P pitch devices requiring \SI{<= 20}{\nm} alignment accuracy between layers, depending on gate design and becoming more stringent for reduced pitch.  Such repeated overlay is nontrivial with e-beam lithography but is well within capability for advanced-node extreme-ultraviolet optical lithography \cite{euvLitho2016}.  With increasing BEOL layer count, crosstalk and other effects arising from BEOL resistance and ILD capacitance should deteriorate performance, though at present we have not experimentally observed signal integrity issues with three layers.  Similarly, we do not experimentally observe any impacts to qubit performance due to increased Johnson noise from the resistances of additional vias, though it may become problematic with higher BEOL layer counts. Future work may include modeling of realistic device stackups and material properties to derive SLEDGE interconnect bandwidth limitations.  Reductions in via/metal resistance, for example with superconducting materials, may mitigate signal degradation, if needed.  Lastly, the sensitivity of quantum-dot-based charge sensors may limit the extension of SLEDGE gate arrays; the sensitivity scales with the distance $r$ between a sensor and a quantum dot as $1/r^{3}$ \cite{PhysRevApplied.6.054013}. Additional peripheral charge sensors are possible, but the sensitivity scaling will make the charge-state tuning of large 2D arrays challenging, though we have not observed difficulty with two rails of plungers and a single sensor. Other measurement schemes may be necessary for significantly larger 2D arrays \cite{Colless2013,West2019}. Aside from effective limitations, the SLEDGE process has compatibility opportunities with other semiconductor platforms used for hosting spin qubits and control schemes, such as Ge/SiGe heterostructures and metal-oxide-semiconductor devices, provided the Ohmic contact process is appropriately modified.

We have developed a technology utilizing the flexible SLEDGE platform in conjuction with a multilayer BEOL process to encode EO DFS qubits in an extendable 2D quantum dot array fabricated on a Si/SiGe heterostructure.
EO qubits are well-suited for 2D arrays as they can be configured to avoid poorly performing quantum dots and low-coherence exchange interaction sites. The calibration of arbitrary EO qubits in SLEDGE 2D spin arrays is straightforwardly accomplished via the calibration of arbitrary exchange axes. Calibrated spin swaps can be used to instantiate qubits at any desired location within a spin array and, in combination with full-permutation dynamical decoupling \cite{SunNZ2024}, for sequential SPAM of multiple EO qubits in 2D arrays. 
EO qubits in 2D arrays may be able to take advantage of full connectivity to achieve encoded two-qubit operations with fewer pulses than in linear arrays \cite{HRL_2QEO,SetiawanPhysRevB2014}, potentially improving EO two-qubit gate fidelities. 
Future work in SLEDGE 2D arrays will evaluate the performance of EO two-qubit gates utilizing such connectivity, as well as simultaneous multiqubit initialization and sequential measurement.

\begin{acknowledgments}
We thank John Carpenter for assistance with all figures.
\end{acknowledgments}

\bibliography{multirail_bib}

\end{document}